\documentclass[useAMS,usenatbib]{mn2e}

\usepackage{subeqnarray,graphicx,amsmath}
\usepackage{multirow}
\newcommand{\MS}{\ifmmode{\,}\else\thinspace\fi{\rm M}\ifmmode_{\odot}\else$_{\odot}$\fi}
\newcommand{\LS}{\ifmmode{\,}\else\thinspace\fi{\rm L}\ifmmode_{\odot}\else$_{\odot}$\fi}
\newcommand{\RS}{\ifmmode{\,}\else\thinspace\fi{\rm R}\ifmmode_{\odot}\else$_{\odot}$\fi}
\newcommand{\Ke}{\ifmmode{\,}\else\thinspace\fi{\rm K}}
\newcommand{\iu}{{\rm i}}
\newcommand{\teff}{\ifmmode T_{\rm eff}\else$T_{\rm eff}$\fi}
\title[Period doubling and Blazhko modulation in BL~Herculis hydrodynamic models]{Period doubling and Blazhko modulation in BL~Herculis hydrodynamic models.}
\author[R. Smolec \& P. Moskalik]
{R. Smolec\thanks{E-mail:
smolec@camk.edu.pl} and
P. Moskalik\\
Nicolaus Copernicus Astronomical Centre, ul. Bartycka 18, 00-716 Warszawa, Poland\\
}

\begin{document}

\date{Accepted . Received ; in original form }

\pagerange{\pageref{firstpage}--\pageref{lastpage}} \pubyear{2012}

\maketitle

\label{firstpage}

\begin{abstract}
We present the hydrodynamic BL~Herculis-type models which display a long-term modulation of pulsation amplitudes and phases. The modulation is either strictly periodic or it is quasi-periodic, with the modulation period and modulation pattern varying from one cycle to the other. Such behaviour has not been observed in any BL~Her variable so far, however, it is a common property of their lower luminosity siblings -- RR~Lyrae variables showing the Blazhko effect. These models provide a support for the recent mechanism proposed by Buchler \& Koll\'ath to explain this still mysterious phenomenon. In their model, a half-integer resonance that causes the period doubling effect, discovered recently in the Blazhko RR~Lyrae stars, is responsible for the modulation of the pulsation as well. Although our models are more luminous than is appropriate for RR~Lyrae stars, they clearly demonstrate, through direct hydrodynamic computation, that the mechanism can indeed be operational. 

Of great importance are models which show quasi-periodic modulation -- a phenomenon observed in Blazhko RR~Lyrae stars. Our models coupled with the analysis of the amplitude equations show that such behaviour may be caused by the dynamical evolution occurring in the close proximity of the unstable single periodic saddle point.
\end{abstract}


\section{Introduction}\label{sec.intro}

Nonlinear modelling is one of the key methods to study the large amplitude variability of classical pulsators: RR~Lyrae stars and Cepheids. One of the notable successes of nonlinear pulsation theory is the explanation of resonant effects in classical pulsators. One of the effects is the bump progression in the light/radial velocity curves of classical Cepheids pulsating in the fundamental mode (so-called Hertzsprung progression). The effect is caused by the 2:1 resonance between the fundamental mode and the second overtone, the latter mode being resonantly excited to high amplitude \citep[e.g.][]{ss76,kovb89,bmk90,bms00}. Due to resonant nonlinear phase synchronisation the second overtone is not visible separately, but manifests itself as a distortion in the light/radial velocity curve. A similar effect is present in classical Cepheids pulsating in the first overtone \citep[see e.g.][]{kienzle,fbk00}. Another interesting resonant effect, which we understand thanks to nonlinear pulsation theory, is a period-doubling behaviour -- alternating deep and shallow minima in the light/radial velocity curves. It is a characteristic feature of RV~Tau variables, a group of pulsators often considered to be a subgroup of type-II Cepheids \citep[e.g.][]{wallerstein,szabados}. The period doubling effect is caused by the half-integer resonances, as analysed by \cite{mb90}. In case of RV~Tau variables, the 5:2 resonance between the fundamental mode and the second overtone is crucial. Period doubling was also discovered in Blazhko RR~Lyrae stars observed with the satellite mission {\it Kepler} \citep{kol10, szabo10} and explained by \cite{kms11} as a result of a 9:2 resonance between the fundamental mode and the 9th overtone. Recently \cite{sosz_cep_blg} and \cite{ssm12} reported the discovery of the period doubling effect in a BL~Her star, demonstrating the predictive power of the nonlinear pulsation theory. Existence of period-doubled BL~Her stars was in fact predicted twenty years earlier by \cite{bm92} \citep[see also][]{bb94}, who found the effect in their radiative hydrodynamic models. \cite{ssm12} confirmed that the 3:2 resonance between the fundamental mode and the first overtone is the cause of the alternations, as analysed earlier by \cite{bm92}.

In \cite{ssm12} we have studied the period doubling effect in BL~Her models with our state-of-the-art convective hydrocodes \citep{sm08a}. During the test computations with the decreased eddy viscosity we identified a class of models showing the modulation of pulsation amplitude and phase, which is either strictly periodic or quasi-periodic. Such behaviour has not been detected in any BL~Her star so far. Also, because of the strongly reduced eddy-viscosity, the pulsations are violent and spurious spikes appear in the light curves. Still, the models are of great importance for the lower luminosity cousins of BL~Her stars -- RR~Lyrae pulsators, in which the more or less periodic modulation of pulsation amplitude and phase -- the Blazhko effect -- is a common property \citep[e.g.][]{kk08}. 

The amplitude modulation in RR~Lyrae stars is one of the most disturbing problems of stellar astrophysics. Although it was discovered more than century ago \citep{blazhko}, its origin is still mysterious. During the recent years extensive ground-based observation campaigns \citep[e.g.][]{kol06,jj09} and a top quality and nearly continuous observations of space missions {\it CoRoT} and {\it Kepler} allowed to rule out the two models proposed to explain the Blazhko effect, the magnetic oblique rotator/pulsator model and the non-radial resonant rotator/pulsator model \citep[see][for a review]{GezaSF}. One of the main drawbacks of these models is the predicted clock-work modulation, while it became clear, with the advent of satellite data, that the Blazhko effect can be a very irregular phenomenon \citep[see e.g.][]{eg11,eg12,kol11}. The recent idea of \cite{st06}, which assumes that modulation of turbulent convection by transient magnetic fields causes the modulation of pulsation was also questioned \citep{smk11,mks12}. The discovery of period doubling effect in some Blazhko variables led to the new model behind the Blazhko modulation -- the radial mode resonance model proposed by \cite{bk11}. Using the amplitude equations' formalism \citep[AE, see e.g.][]{bg84} \cite{bk11} showed that the same resonance that causes the period doubling in Blazhko variables, i.e. the 9:2 resonance between the fundamental mode and the ninth overtone, can also cause either periodic or chaotic modulation of the pulsation. The AE formalism is a powerful tool to study the nonlinear pulsation, specifically the possible limit cycles and their stability. The nature of the computed limit cycles, however, depends on the arbitrarily assumed values of the saturation and resonant coupling coefficients, for which realistic values are too difficult to compute \citep[see e.g.][]{klapp,nowakowski}. Consequently, it is not clear whether the specific solution of the AEs in which pulsation is modulated, can be represented in the real stars. This must be confirmed with realistic nonlinear hydrodynamic models. Attempts to find modulation of pulsation in hydrodynamic models of RR~Lyrae stars did not lead to success so far, likely because of the surface nature of the involved 9th order overtone, which is difficult to model \citep{bk11}. In this paper we present the hydrodynamic models of BL~Her variables, more luminous, but otherwise (masses, chemical composition, evolution history) siblings of RR~Lyrae stars. In these models a low order 3:2 resonance, between the fundamental mode and the first overtone causes the modulation of pulsation. Our calculations demonstrate for the first time that the mechanism proposed by \cite{bk11} can be indeed operational in hydrodynamical models. In addition, they illustrate how the quasi-periodic modulation may arise. Thus, they provide a strong support for the radial mode resonance model of the Blazhko effect.

We first present our hydrodynamic models which display modulation of pulsation (Section~\ref{sec.hydro}) and next we demonstrate that the computed behaviour can be captured and understood with the AE formalism (Section~\ref{sec.ae}). In Section~\ref{sec.concl} we comment on the reliability of the models and mechanisms responsible for the wealth of detected behaviours. Finally, in Section~\ref{sec.impl} we discuss the implications for understanding the Blazhko effect in RR Lyrae stars.

\section{Hydrodynamic models}\label{sec.hydro}

\begin{figure}
\centering
\resizebox{\hsize}{!}{\includegraphics{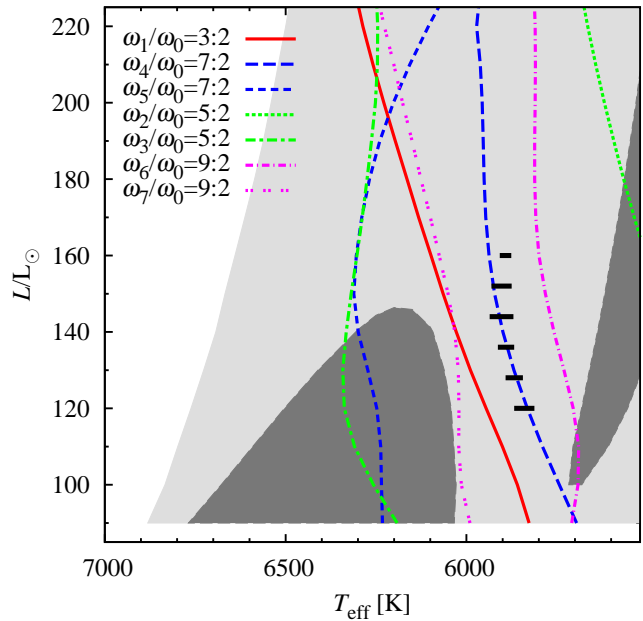}}
\caption{The HR diagram for the computed models of $M=0.55\MS$ and $Z=0.0001$. Instability strips for the fundamental mode and the first overtone are indicated with the light- and dark-shaded areas respectively. Loci of several resonances are plotted as indicated in the key. Horizontal line segments show the domain of modulation of the pulsations.}
\label{fig.hyd.hr}
\end{figure}

The hydrodynamic models were computed with the Warsaw nonlinear, convective, pulsation codes \citep{sm08a}. Numerical parameters (zoning) and convective parameters, except of eddy viscosity, are the same as in \cite{ssm12} (see their section~3.1 and table~2, set P1). The eddy viscosity ($\alpha_m$ parameter) was decreased from $0.25$ to $0.05$. The resulting fundamental mode instability strip (IS) -- see Fig.~\ref{fig.hyd.hr} -- is very wide, the red edge was not reached in our computations. For the first overtone we observe an additional thin instability strip extending to the red of the usual first overtone IS and towards the higher luminosities. Consequently, our models do not present a realistic scenario for BL~Her pulsation (see Section~\ref{sec.concl} for detailed discussion). Yet, they show a very interesting feature, namely the periodic amplitude modulation on top of the period doubling alternations -- see bottom panel of Fig.~\ref{fig.hyd.lc} for an example. In Fig.~\ref{fig.hyd.hr} we show the location of these interesting models in the HR diagram (horizontal line segments). Over-plotted are loci of several half-integer resonances. Each of them can, in principle, cause the period doubling effect and modulation of pulsation \citep[][Section~\ref{sec.ae}]{mb90}. Although the modulation domain coincides with the loci of the 7:2 resonance with the 4th overtone, the modulation of pulsation is most likely caused by the 3:2 resonance with the first overtone, which is responsible for the period doubling effect in BL~Her models \citep{ssm12}. The origin of the modulation will be discussed in more detail in Section~\ref{sec.concl}.

Because of strongly decreased eddy viscosity the models are numerically demanding. Violent pulsation require very short time steps leading to extremely slow integration. Consequently, only a limited nonlinear model survey was conducted.  All models have $M=0.55\MS$ and $Z=0.0001$ ($X=0.76$). We used the OP opacities \citep{sea05} and adopted a solar mixture of \cite{a04}. The nonlinear models were computed along six lines of constant luminosity, $L\in\{120\LS,\ 128\LS,\ 136\LS,\ 144\LS,\ 152\LS,\ 160\LS\}$ and cover a limited range in effective temperatures -- large enough to determine the boundaries of the domain in which modulation of pulsation takes place.  We first describe the model sequences in which all models show strictly periodic modulation of pulsation (Section~\ref{ssec.periodic}). Next we describe the model sequences along which models with quasi-periodic modulation were found (Section~\ref{ssec.chaotic}).

\subsection{Models with strictly periodic modulation of pulsation}\label{ssec.periodic}

\begin{figure}
\centering
\resizebox{\hsize}{!}{\includegraphics{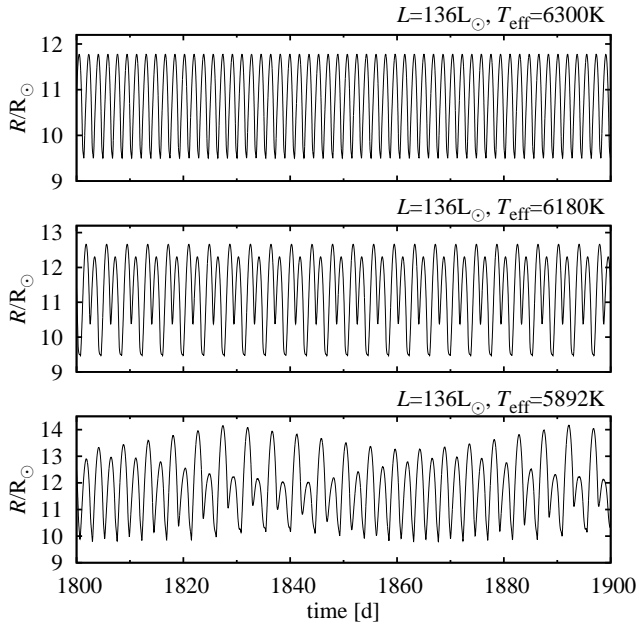}}
\caption{Radius variation for the three models of $136\LS$ and different effective temperatures.}
\label{fig.hyd.lc}
\end{figure}

Strictly periodic modulation was found in all models along four sequences of highest luminosity (see Fig.~\ref{fig.hyd.hr}). The pulsation scenario is discussed based on the model sequence of $L=136\LS$. The radius variation for the three models of different effective temperatures is presented in Fig.~\ref{fig.hyd.lc}. The hottest model ($T_{\rm eff}=6300$\thinspace K, top panel of Fig.~\ref{fig.hyd.lc}) displays a single-periodic, fundamental mode pulsation. As we decrease the model's effective temperature the period-doubling domain emerges. The radius variation is illustrated in the middle panel of Fig.~\ref{fig.hyd.lc} for model of $T_{\rm eff}=6180$\thinspace K. Alternation of radii maxima and minima is clear. At still lower effective temperatures, periodic modulation of pulsation on top of the period doubling effect is found in the models -- bottom panel of Fig.~\ref{fig.hyd.lc}. The domain in which we detect this behaviour is not large, its width is $\approx 50$\thinspace K (see Fig.~\ref{fig.hyd.hr}). To the red of this domain the period doubling only (no modulation) behaviour is present again, and next we find a single-periodic pulsation.

\begin{figure}
\centering
\resizebox{\hsize}{!}{\includegraphics{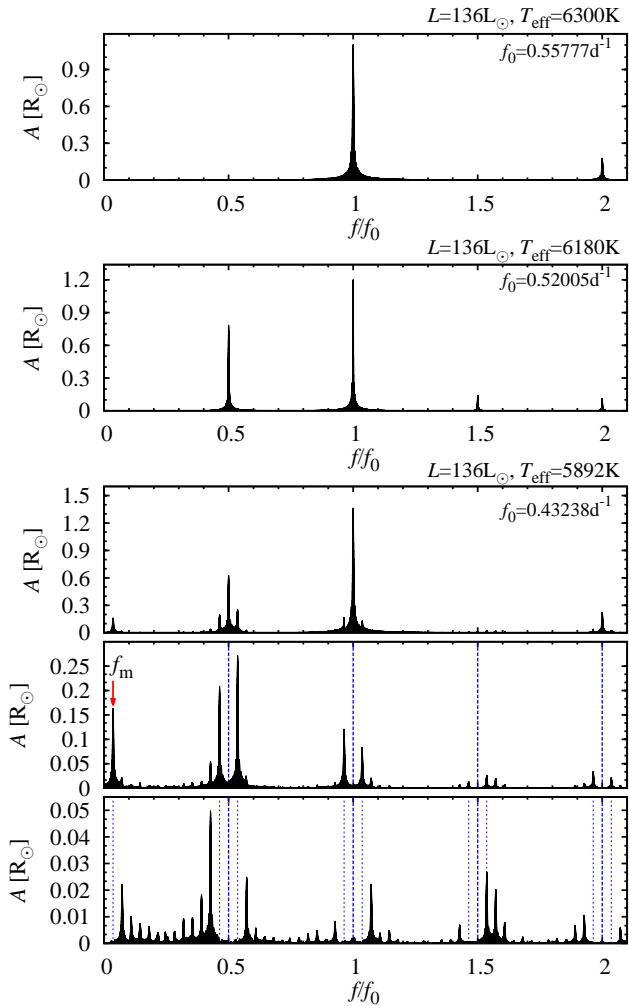}}
\caption{Frequency spectra for the models of Fig.~\ref{fig.hyd.lc}. {\it Top panel:} frequency spectrum for model of $T_{\rm eff}=6300$\thinspace K that exhibits single periodic pulsation; {\it middle panel:}  frequency spectrum for model of $T_{\rm eff}=6180$\thinspace K that exhibits period doubling effect; {\it bottom panel:} Prewhitening sequence for model of $T_{\rm eff}=5892$\thinspace K with periodic modulation of pulsation. Removed frequencies are marked with vertical dashed lines}.
\label{fig.hyd.fs1}
\end{figure}

\begin{figure*}
\centering
\resizebox{\hsize}{!}{\includegraphics{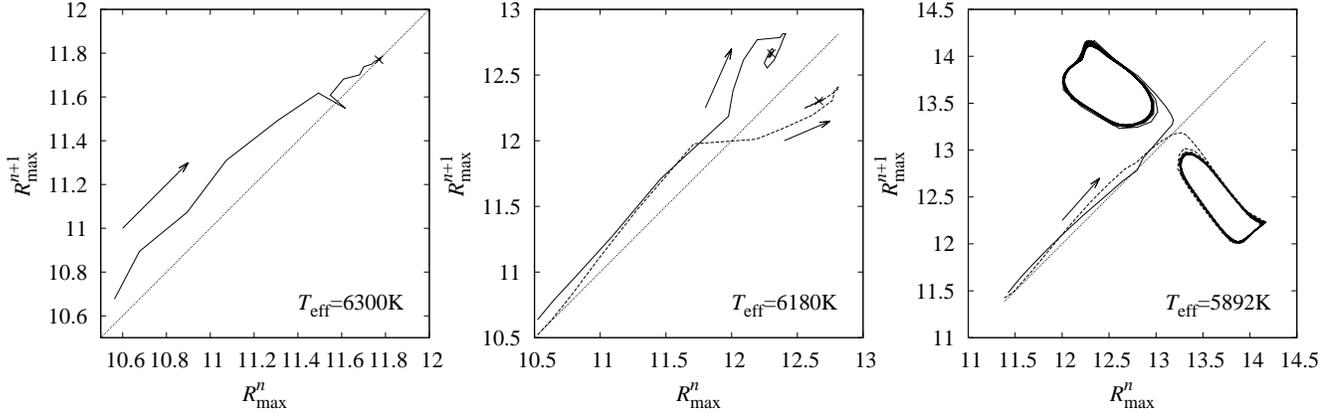}}
\caption{First return maps for radii maxima (in units of \RS) for models of Fig.~\ref{fig.hyd.lc}.}
\label{fig.hyd.rm}
\end{figure*}

The described behaviour is also clear from the analysis of the frequency spectra of the three models of Fig.~\ref{fig.hyd.lc}, presented in Fig.~\ref{fig.hyd.fs1}. For better readability the frequencies are normalised with the fundamental mode frequency, $f_0$. In the top two panels of Fig.~\ref{fig.hyd.fs1} we show the frequency spectra for the two hotter models for which no modulation was detected. For single periodic pulsation in the model of $T_{\rm eff}=6300$\thinspace K (top panel of Fig.~\ref{fig.hyd.fs1}) only the fundamental mode and its harmonics are present in the frequency spectrum. For the cooler model of $T_{\rm eff}=6180$\thinspace K (middle panel of Fig.~\ref{fig.hyd.fs1}) strong signal at the half-integer frequencies (sub-harmonics) is detected as well. It is a characteristic signature of a period doubling effect. After prewhitening with the fundamental mode and its harmonics ($T_{\rm eff}=6300$\thinspace K) and sub-harmonics ($T_{\rm eff}=6180$\thinspace K) we find no additional signal in the data. In the bottom panel of Fig.~\ref{fig.hyd.fs1} we show the prewhitening sequence for the model of $T_{\rm eff}=5892$\thinspace K, that shows the modulation of pulsation on top of period doubling (see bottom panel of Fig.~\ref{fig.hyd.lc}). The top panel in the prewhitening sequence shows the frequency spectrum of the original hydrodynamic time-series. The fundamental mode, its harmonic and two sub-harmonics are well visible. At $f_0$, $1/2f_0$ and at low frequencies, additional peaks are also present, which become clearly visible after prewhitening with the fundamental mode its harmonics and subharmonics (middle panel in the prewhitening sequence). These peaks correspond to the modulation of pulsation. With the removed frequencies they form the equidistant multiplets. The separation between the multiplet components corresponds to the modulation frequency, $f_{\rm m}$, which is also detected independently in the frequency spectrum. For the discussed case the modulation period is $P_{\rm m}=1/f_{\rm m}=63.8$\thinspace days, which corresponds to $27.6$ fundamental mode pulsation cycles.

The modulation of pulsation can also be illustrated through construction of the return maps. In Fig.~\ref{fig.hyd.rm} we show the first return map for the radii maxima, i.e., a plot of $R_{\rm max}^{n+1}$ vs. $R_{\rm max}^{n}$, where $n$ counts the pulsation cycles, for the three discussed models. The return maps are plotted for the full model integration, including the initial transient phase. For single periodic model of $T_{\rm eff}=6300$\thinspace K, after initial transient, the consecutive radii maxima fall at a single point located on the diagonal (cross in the leftmost panel of Fig.~\ref{fig.hyd.rm}). It corresponds to a fundamental mode limit cycle pulsation. For models of $T_{\rm eff}=6180$\thinspace K and $T_{\rm eff}=5892$\thinspace K, which show the period doubling effect, it is convenient to connect each second pair of radii maxima (middle and right panel of Fig.~\ref{fig.hyd.rm}). After initial growth of the amplitude of single-periodic pulsation the so constructed curves diverge as a result of the raise of the period doubling effect. At the limit cycle pulsation the return maps for the two models are different. For the model of $T_{\rm eff}=6180$\thinspace K, that shows the period doubling effect and no modulation (middle panel of Fig.~\ref{fig.hyd.rm}), the consecutive radii maxima alternate between two points, located symmetrically with respect to the diagonal. For the model of $T_{\rm eff}=5892$\thinspace K that shows the periodic modulation of pulsation (right panel of Fig.~\ref{fig.hyd.rm}), each second pair of radii maxima follow a route along a separate loop. The modulation is strictly periodic, each loop can be described by a single curve -- the broadening visible in Fig.~\ref{fig.hyd.rm} is caused by connecting with straight line segments.

\begin{figure}
\centering
\resizebox{\hsize}{!}{\includegraphics{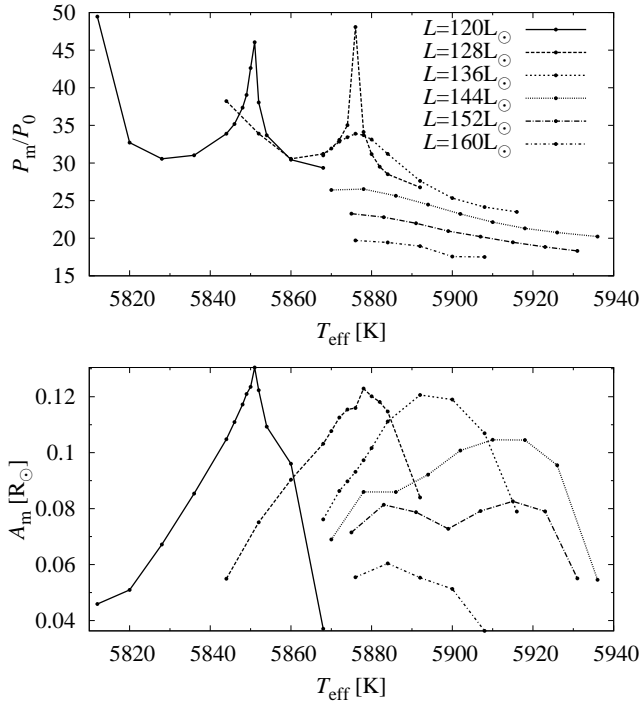}}
\caption{Modulation periods in units of the fundamental mode pulsation period (upper panel) and modulation amplitudes (bottom panel) as a function of effective temperature and luminosity (different line types).}
\label{fig.hyd.mp}
\end{figure}

The described properties of modulation are qualitatively the same for model sequences of higher luminosity, i.e. for $L\in \{144\LS,\ 152\LS,\ 160\LS\}$. In the upper panel of Fig.~\ref{fig.hyd.mp} we plot the modulation periods as a function of model parameters: luminosity and effective temperature. In the bottom panel the corresponding modulation amplitude, i.e. the amplitude of the highest modulation peak at the vicinity of $f_0$ (either $f_0-f_{\rm m}$ or $f_0+f_{\rm m}$) is plotted. The typical modulation periods vary from $\sim 16$ to $\sim 35$ fundamental mode pulsation cycles ($45$ -- $106$ days). In general, the higher the model luminosity and the higher its effective temperature the shorter the modulation period. The modulation amplitudes decrease towards the edges of the domain in which modulation of pulsation was found. This indicates that the appearance of the modulation domain, as the model's effective temperature is changed, is a supercritical bifurcation. Also, for $L\ge 136\LS$, the higher the luminosity the lower the modulation amplitudes. It is clear that the modulation domain shrinks towards higher luminosities and likely does not extend far beyond $160\LS$. 

In case of the least luminous models of $120\LS$ and $128\LS$ sharp features are clearly visible in the progressions of modulation periods and amplitudes with effective temperature. These models are discussed in detail in the next Section. 

\subsection{Models with quasi-periodic modulation of pulsation}\label{ssec.chaotic}

\begin{figure*}
\centering
\resizebox{\hsize}{!}{\includegraphics{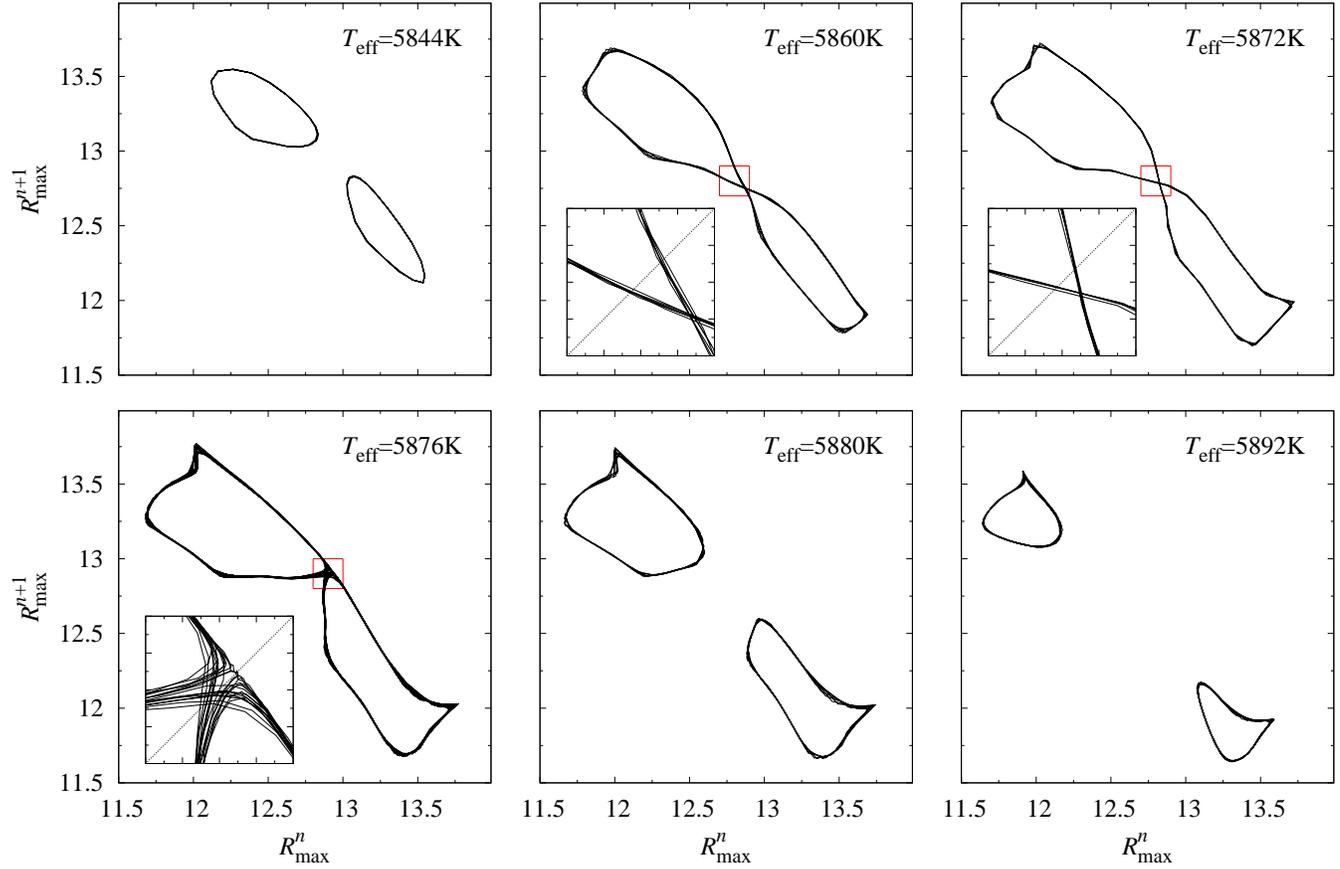}}\\
\caption{First return maps for maximum radii (in units of \RS) for several models of $128\LS$. The insets in three panels show the zoom of the marked regions.}
\label{fig.hyd.128.rm}
\end{figure*}

\begin{figure*}
\centering
\resizebox{\hsize}{!}{\includegraphics{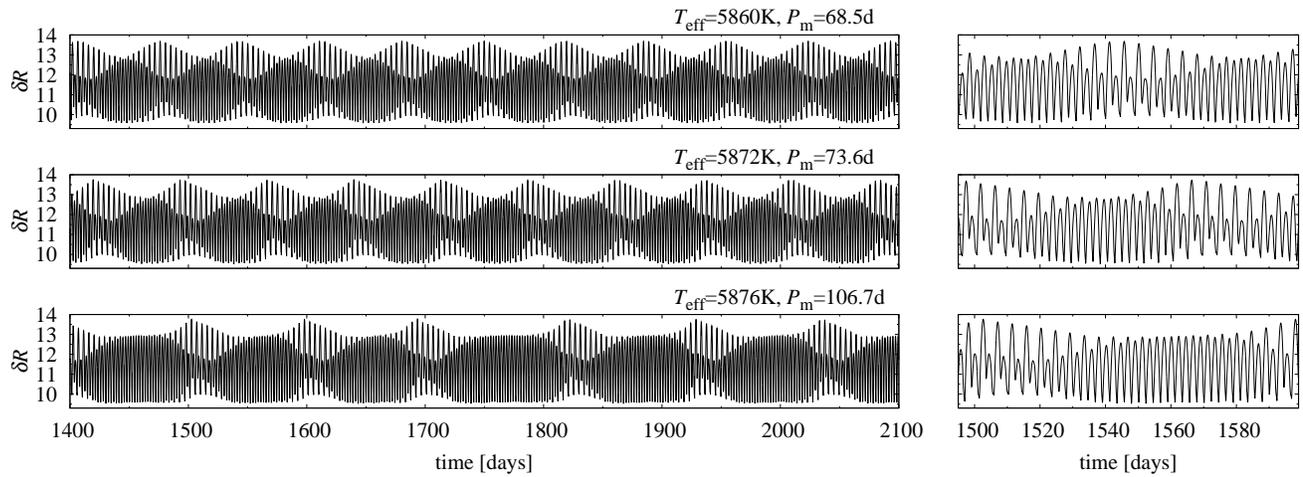}}\\
\caption{Radius variation for the three models of $128\LS$ and different effective temperatures. Right panels show the zoom into a $100$-day segment.}
\label{fig.hyd.128.lc}
\end{figure*}

\begin{figure}
\centering
\resizebox{\hsize}{!}{\includegraphics{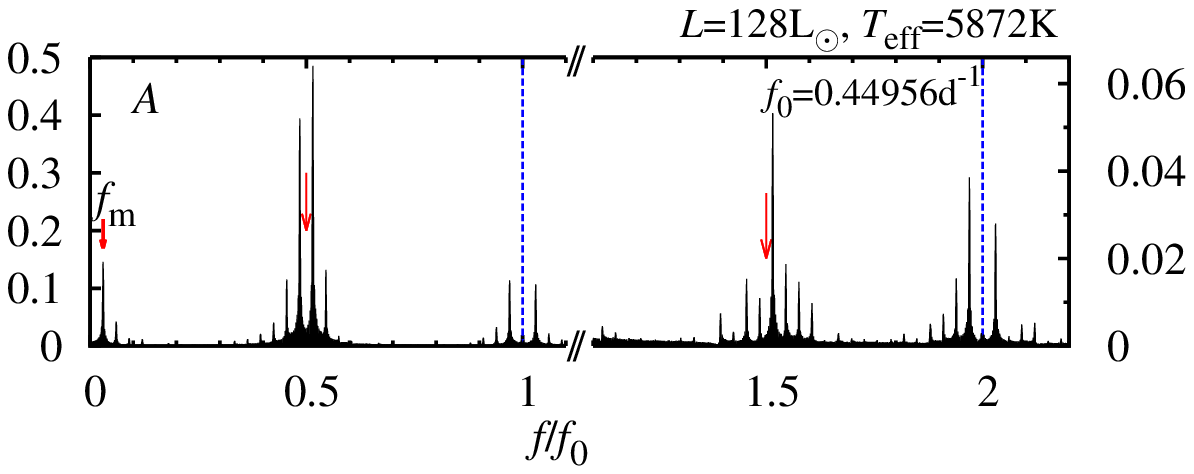}}
\resizebox{\hsize}{!}{\includegraphics{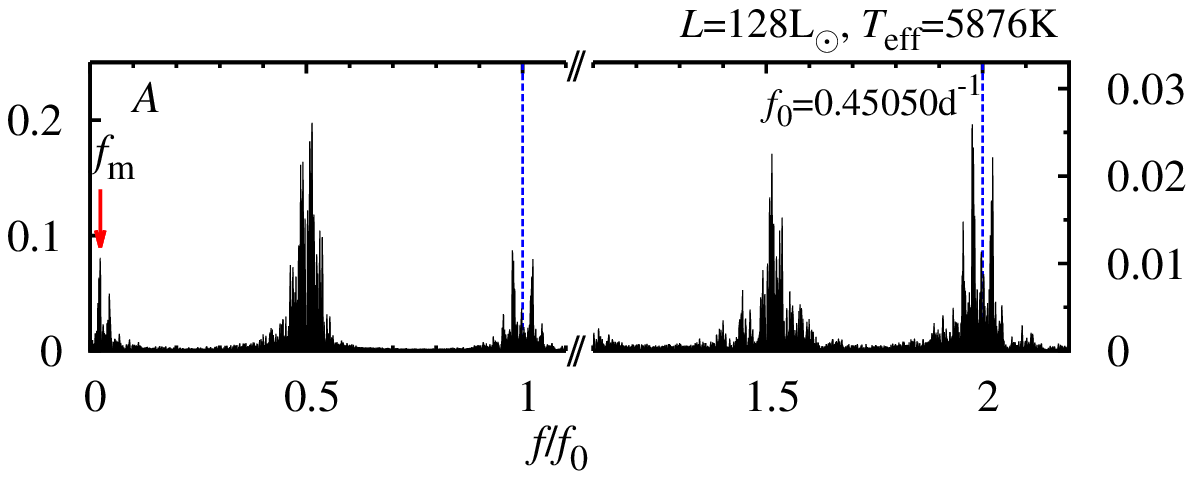}}
\caption{Frequency spectra after prewhitening with the fundamental mode and its harmonics for two models of 128\LS\ sequence. Note different scales on the vertical axis in the left and right sides of the Figure. Removed frequencies are marked with dashed lines, while arrows (top panel) show the exact location of the sub-harmonic frequencies, $1/2f_0$ and $3/2f_0$.}
\label{fig.hyd.587276.fs}
\end{figure}

In Fig.~\ref{fig.hyd.128.rm} we plot the return maps for the limit cycle pulsation (initial transient phase was omitted) for several models along the $128\LS$ sequence. As before, each second pair of points were connected (period doubling). For the coolest and two hottest models the return maps are qualitatively the same as discussed in the previous Section (two separate loops). These models show strictly periodic modulation of pulsation. To explain the sharp increase of the modulation periods visible in Fig.~\ref{fig.hyd.mp}, we focus our attention on the three models with effective temperatures of $5860$\thinspace K, $5872$\thinspace K and $5876$\thinspace K. Their modulation periods increase and are equal to $68.5$\thinspace d, $73.6$\thinspace d and $106.7$\thinspace d, respectively. The radius variation for these models is plotted in Fig.~\ref{fig.hyd.128.lc}. The return maps have a different topology from that discussed so far. They have a shape of figure `8'. For the two models of $5860$\thinspace K and $5872$\thinspace K the pulsation modulation is periodic, but the higher/lower maximum pulsation cycles trade places from one modulation cycle to the other. This behaviour can be followed in Fig.~\ref{fig.hyd.128.lc} (in particular in rightmost panels). Consequently, the pattern of period doubling repeats after two modulation cycles.
 We note that this behaviour was recently found in the {\it Kepler} short cadence data for RR~Lyrae (Moln\'ar et al., submitted to MNRAS), although the swaping of higher/lower maximum cycles is very irregular. The frequency spectrum for the model of $T_{\rm eff}=5872$\thinspace K (it is qualitatively the same for the model of $T_{\rm eff}=5860$\thinspace K), after prewhitening with the fundamental mode and its harmonics, is plotted in the top panel of Fig.~\ref{fig.hyd.587276.fs}. At the position of the sub-harmonics, i.e., exactly at $1/2f_0$ and $3/2f_0$, we detect no signal. Instead a comb of equidistant frequencies, with peak separation corresponding to $f_{\rm m}$ is centred exactly on the location of the missing sub-harmonics. For $f_0$ and the harmonic frequencies (i.e. $2f_0$, $3f_0$, etc.) we observe a multiplet structure of the same properties as discussed earlier (see Fig.~\ref{fig.hyd.fs1}).  

Even more complicated behaviour is observed for the model of $T_{\rm eff}=5876$\thinspace K. We note that this is the model of the longest, peak modulation period (Fig.~\ref{fig.hyd.mp}). In fact the period of the modulation cycle, if measured from one epoch of maximum pulsation amplitude to the other, may strongly vary for this model which is well visible in the bottom panel of Fig.~\ref{fig.hyd.128.lc}. For the plotted cycles we estimate it to be $96$\thinspace d, $95$\thinspace d, $129$\thinspace d, $106$\thinspace d and $107$\thinspace d. The modulation is quasi-periodic. The first return map for this model is presented in Fig.~\ref{fig.hyd.128.rm}. Analysis of the inset shows that the radii maxima either follow the route along the figure `8'  or follow only its one part. As long as we continued the model integration ($20$\thinspace $000$ fundamental mode pulsation cycles, i.e., more than $400$ modulation cycles) no regularity could be spotted in the described behaviour. We also note that consecutive radii maxima do not follow a single curve, which is well visible in the respective inset in Fig.~\ref{fig.hyd.128.rm}. Consequently the modulation pattern varies from one modulation cycle to the other. This quasi-periodic behaviour is also detected in the frequency spectrum, which is plotted in the bottom panel of Fig.~\ref{fig.hyd.587276.fs}. At $f_0$ and its harmonic we detect the modulation multiplet with the separation corresponding to $f_m$. The inverse,$f_{\rm m}^{-1}$, describes the mean modulation period, which is equal to $106.7$\thinspace d. At the sub-harmonic frequencies broad bands of peaks with no clear structure are visible. These peaks are not coherent just as are the modulation peaks around $f_0$ and its harmonics, clearly indicating the chaotic nature of the model. Still the modulation can be described as more or less periodic, as it is the case in several Blazhko RR~Lyrae stars, with strongly differing modulation cycles \citep[e.g.][]{eg11}. We note that properties of Fourier spectrum for the discussed model closely resemble the properties of Fourier spectra for Blazhko RR Lyrae stars showing the period doubling effect. For these stars bunches of peaks are present at half-integer frequencies indicating the irregular nature of period doubling phenomenon \citep[][]{szabo10,kol11}.

As the models' temperature is increased beyond $5876$\thinspace K, the figure `8' is broken into two separate loops (Fig.~\ref{fig.hyd.128.rm}). Strictly periodic modulation, as described in Section~\ref{ssec.periodic}, is observed again.
 
Analysis of the radius variation and return maps for our models provide a hint why the modulation period varies for these models (as depicted in Fig.~\ref{fig.hyd.mp}) and why the modulation may become quasi-periodic. We focus attention on the three previously discussed models (of \teff=5860\thinspace K, 5872\thinspace K and 5876\thinspace K), for which the radius variation is plotted in Fig.~\ref{fig.hyd.128.lc}. During the modulation the amplitudes of the excited modes, i.e. of the fundamental mode ($A(t)$ in the following), and resonantly coupled first overtone ($B(t)$) vary. For the three models we observe that the period doubling  is quenched at some phases of the modulation cycle. It is visible in the return maps (radii maxima fall close to the diagonal, see insets in Fig.~\ref{fig.hyd.128.rm}) and clearly in Fig.~\ref{fig.hyd.128.lc}. The amplitude of the overtone mode must be very small then. We conclude that the model evolves in direct proximity of the point corresponding to unstable single periodic pulsation, i.e. a point of $A={\rm const}$ and $B=0$. This unstable limit cycle corresponds to a single point on the diagonal of the return map. The topology of return map for \teff=5876\thinspace K (Fig.~\ref{fig.hyd.128.rm}) suggest that it is a saddle point. The exact location of the unstable single periodic limit cycle could be determined with the help of the relaxation technique \citep{stel74}. Unfortunately it is not implemented in our code. Nevertheless, in the next Section, with the help of the amplitude equations formalism, we show that indeed the observed properties of the modulation result from the dynamical evolution close to the unstable saddle point. We will show that at a saddle point the trajectories describing the system evolution are at first attracted, but repelled later on, and that evolution is very slow (the closer to saddle point, the slower the system's evolution). Thus, the models evolving closer to the unstable saddle point have much longer modulation periods (and a much longer phase of suppressed period doubling). In addition once the trajectories evolve very close to the saddle, the system becomes chaotic. When the system evolves far from the saddle point the modulation is strictly periodic and occurs on a shorter time scale.

\begin{figure*}
\centering
\resizebox{\hsize}{!}{\includegraphics{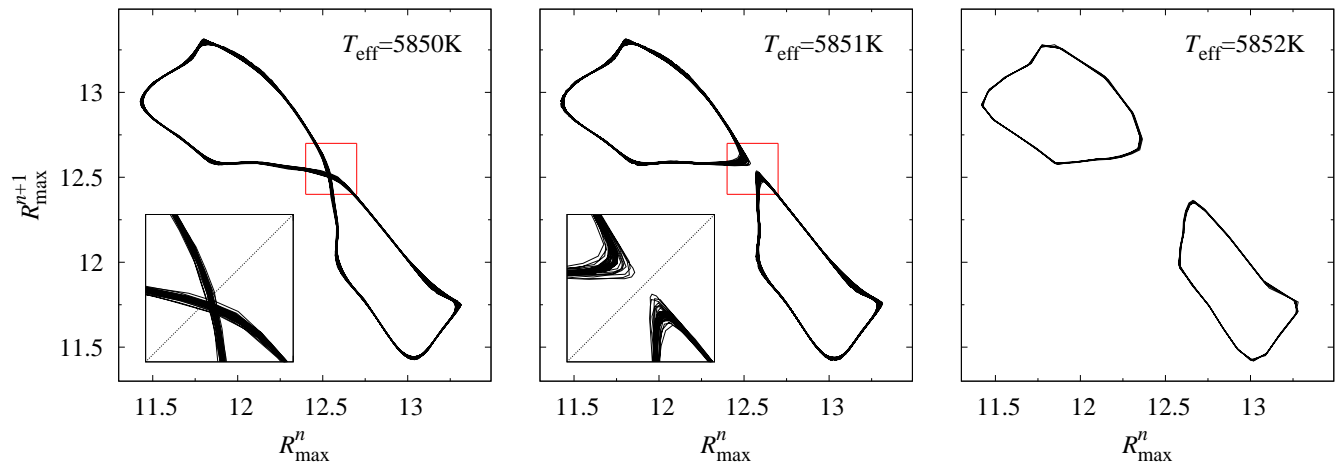}}\\
\caption{First return maps for maximum radii (in units of \RS) for three models of $120\LS$. The insets in the first two panels show the zoom of the marked regions.}
\label{fig.hyd.120.rm}
\end{figure*}

The described properties of radius variation are qualitatively the same for our least luminous model sequence ($120\LS$). In Fig.~\ref{fig.hyd.120.rm} we plot the return maps for three models located at the peak of modulation periods (Fig.~\ref{fig.hyd.mp}). Again we observe that the radii maxima fall close to the diagonal along a broad band (see insets for models of $5850$\thinspace K and $5851$\thinspace K). For the model of $T_{\rm eff}=5851$\thinspace K the radii maxima may follow a route along the loops which approach the diagonal (and the presumed saddle point) at different distances. Corresponding modulation cycles have different length.

In Fig.~\ref{fig.hyd.mp} we also note a rapid increase of the modulation periods at the cool side of the modulation domain, particularly pronounced for the $L=120\LS$ models. This increase again can be explained by the models' evolution close to the unstable saddle point.

Finally we note that also for $136\LS$-sequence the increase of modulation period is visible (Fig.~\ref{fig.hyd.mp}) which is, however, not as prominently marked as in case of the lower luminosity models. 

We now turn to a discussion of the described modulation properties using the amplitude equations' formalism.

\section{Amplitude equations}\label{sec.ae}

The dynamical behaviour we have found in our models can be well captured with the amplitude equations formalism. We follow the analysis of \cite{bk11}, who considered the 9:2 ($9\omega_0=2\omega_9$) resonance responsible for the period doubling detected recently in RR~Lyrae stars \citep{kol10, szabo10, kms11}. All these stars show the Blazhko effect, more or less periodic modulation of the pulsation amplitude. \cite{bk11} showed that the same resonance is able to cause the amplitude modulation. They conducted the parametric study of the appropriate AEs and besides the fixed points they identified two types of oscillatory solutions, either strictly periodic or irregular. Here we follow their approach to study the 3:2 resonance between the fundamental mode and the first overtone which is responsible for the period doubling effect in BL~Her hydrodynamic models \citep{ssm12}. As we noted in Section ~\ref{ssec.periodic}, the stable period two cycle loses its stability and gives rise to a stable cycle in which pulsation is modulated, either periodically or in a quasi-periodic manner. Below we show that this behaviour can be captured with the amplitude equations.

The appropriate set of complex amplitude equations was given by \cite{mb90} \citep[see also][]{mb93b}:
\begin{subeqnarray}
\frac{da}{dt}&=&\big(\iu\omega_a+\kappa_a+Q_a|a|^2+T_a|b|^2\big)a+c_aa^{*2}b^2\,,\\
\frac{db}{dt}&=&\big(\iu\omega_b+\kappa_b+Q_b|b|^2+T_b|a|^2\big)b+c_ba^{3}b^*\,.
\label{eq.cae}
\end{subeqnarray}
The non-resonant part is truncated at the third order. $Q_k$ and $T_k$ are cubic, self- and cross-saturation coefficients, respectively. The last terms on the r.h.s. of \eqref{eq.cae} are resonant coupling terms, with resonant coupling coefficients $c_a$ and $c_b$. These, together with complex amplitudes, are represented as $a=A{\rm e}^{\iu\phi_a}$, $b=B{\rm e}^{\iu\phi_b}$, $c_a=C_a{\rm e}^{\iu\delta_a}$ and $c_b=C_b{\rm e}^{\iu\delta_b}$, which allows to write the above AEs as set of three first order differential equations with real coefficients:
\begin{subeqnarray}
\frac{dA}{dt}&=&\big(\kappa_a+\Re(Q_a)A^2+\Re(T_a)B^2\big)A+\\
&&+C_aA^2B^2\cos(\Gamma+\delta_a)\,,\nonumber\\
\frac{dB}{dt}&=&\big(\kappa_b+\Re(Q_b)B^2+\Re(T_b)A^2\big)B+\\
&&+C_bA^3B\cos(\Gamma-\delta_b)\,,\nonumber\\
\frac{d\Gamma}{dt}&=&2\Delta+\big(2\Im(T_b)-3\Im(Q_a)\big)A^2+\\
&&+\big(2\Im(Q_b)-3\Im(T_a)\big)B^2+\nonumber\\
&&-3C_aAB^2\sin(\Gamma+\delta_a)-2C_bA^3\sin(\Gamma-\delta_b)\,.\nonumber
\label{eq.rae}
\end{subeqnarray}
Above $\Gamma$ is a combination of mode phases, $\Gamma=2\phi_b-3\phi_a$, and $\Delta$ is a frequency mismatch parameter, 
\begin{equation}
\Delta=\omega_b-3\omega_a/2\,.
\label{eq.delta}
\end{equation}
Exact analytical solutions of amplitude equations are possible only for strongly simplified cases. \cite{mb90} considered the AEs for half-integer resonances between the fundamental mode and the higher-order overtones, $(2n+1)\omega_0=2\omega_k$, neglecting the cross-saturation and resonant coupling term  in the equation for the dominating fundamental mode (i.e., $T_a=0$ and $c_a=0$), as well as neglecting the imaginary parts of the cubic coupling coefficients. Such system can have two non-trivial fixed points ($dA/dt=dB/dt=d\Gamma/dt=0$), either single-mode fixed point with non-zero amplitude of the fundamental mode only ($A=\sqrt{-\kappa_a/\Re(Q_a)}$ and $B=0$) or two-mode fixed point with non-vanishing amplitudes of both modes (that in general have to be derived numerically). Because of the half-integer resonance, such solution corresponds to a period-doubled limit cycle. Oscillatory solutions for the amplitudes are possible as well and were found during the numerical integration of eqs.~(\ref{eq.rae}). To simplify the equations we followed \cite{bk11} and neglected the imaginary parts of the cubic saturation coefficients. Periodic modulation of amplitudes, corresponding to hydrodynamic models described in Section~\ref{sec.hydro} was relatively easy to find. We adopted the following set of the saturation and coupling coefficients: $\kappa_a=1.0$, $\kappa_b=-0.01$, $\Re(Q_a)=1$, $\Re(Q_b)=1.1$, $\Re(T_a)=5.0$, $\Re(T_b)=1.0$, $C_a=5.75$, $C_b=4.0$, $\delta_a=1.95$, $\delta_b=1.0$ and integrated the AEs with different mismatch parameter $\Delta$. These models correspond to our hydrodynamic models of different effective temperatures. The radius variation was reconstructed according to \citep{mb90}
\begin{equation}
\delta R(t)=A{\rm e}^{\iu\omega_at}+B{\rm e}^{\iu(\Gamma/2)}{\rm e}^{\iu(3/2)\omega_at}+{\rm h.o.t.}\,,
\label{eq.rad}
\end{equation}
with higher order terms (${\rm h.o.t.}$) neglected. For the pulsation frequency, $\omega_a$,  we choose $\omega_a=2\pi/P_a=2\pi/0.1$ so that the relative growth rate, $\kappa_aP_a=0.1$ is typical for BL~Her models. We note, that we do not expect a 1:1 equivalence between the hydrodynamic models and results of AEs integration, not only because of arbitrarily assumed coefficients of eq.~(\ref{eq.rae}), but also because of neglect of the higher order terms in radii reconstruction (eq.~\ref{eq.rad}).

\begin{figure*}
\centering
\resizebox{\hsize}{!}{\includegraphics{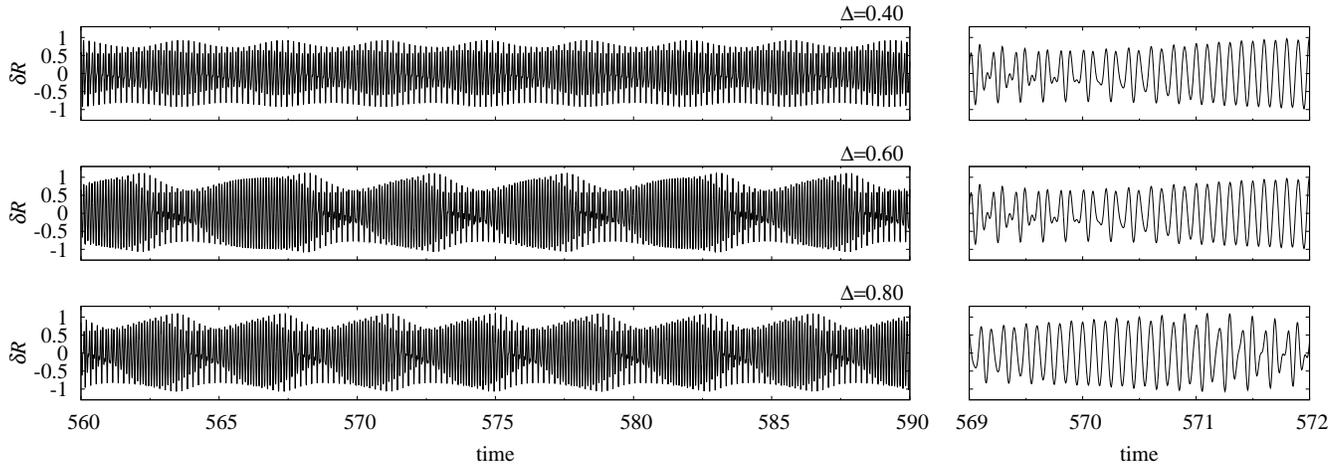}}
\caption{Radius variation for three integrations of AEs with different $\Delta$ parameter. Note the different lengths of consecutive modulation cycles for $\Delta=0.60$.}
\label{fig.ae_lc}
\end{figure*}

\begin{figure*}
\centering
\resizebox{\hsize}{!}{\includegraphics{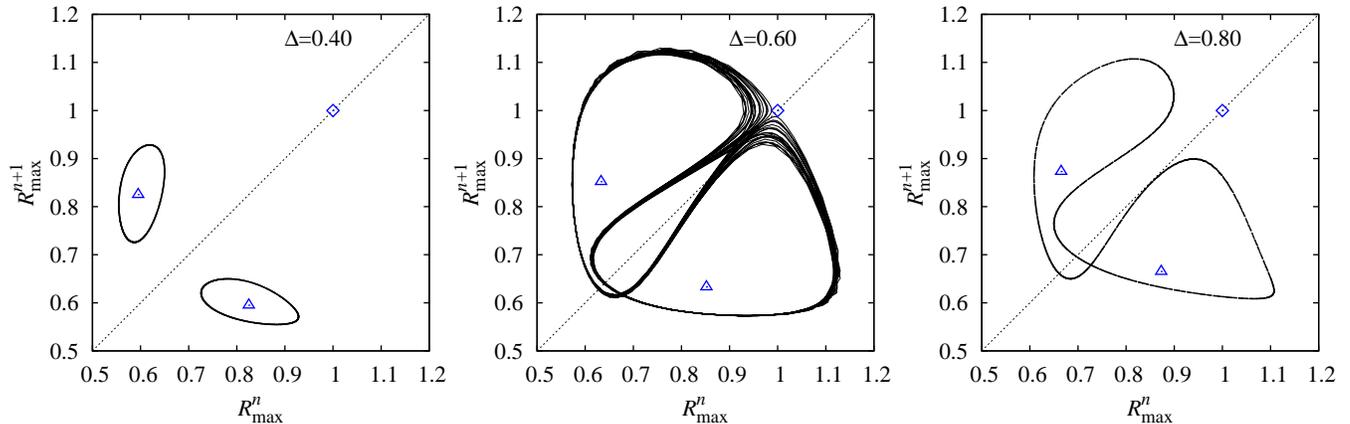}}
\caption{First return maps for radii maxima for models with different $\Delta$. Diamond and triangles mark the location corresponding to unstable single-mode and two-mode fixed points, respectively.}
\label{fig.ae_rm}
\end{figure*}

In Fig.~\ref{fig.ae_lc} we show the radius variation for three different values of $\Delta$. Qualitatively the same behaviour is visible as in the hydrodynamic models (Fig.~\ref{fig.hyd.128.lc}) -- clear amplitude modulation on top of the period doubling behaviour. The modulation is periodic for $\Delta=0.4$ and $\Delta=0.8$, and is clearly quasi-periodic for $\Delta=0.6$\footnote{The apparently large values of the mismatch parameter, $\Delta$, result from the adopted normalization ($\kappa_a=1$).}. The corresponding return maps on successive radii maxima are presented in Fig.~\ref{fig.ae_rm}. Diamond and triangles mark the location of radii maxima corresponding to the single-mode fixed point (diamond) and two-mode fixed point (triangles). The fixed points are unstable as their stability eigenvalues indicate. For the single-mode point the three eigenvalues are all real, one is positive and two are negative, thus the single-mode fixed point is an unstable saddle point. For two-mode fixed point one eigenvalue is real and negative, the other two are a complex conjugate pair with positive real parts. The two-mode fixed point is thus an unstable spiral point. For models with $\Delta=0.40$ and $\Delta=0.80$ the radii maxima fall along single curves encircling the unstable two mode fixed points and far from the unstable single-mode point. These return maps are qualitatively the same as for our hydromodels that display the strictly periodic modulation (e.g. the models of $\teff=5844$\thinspace K or $\teff=5872$\thinspace K in Fig.~\ref{fig.hyd.128.rm}). The behaviour is very different for $\Delta=0.6$. The radii maxima may fall at direct proximity of the unstable single-mode point and may follow a very different routes during the modulation. This is qualitatively the same behaviour as described for models of $T_{\rm eff}=5876$\thinspace K (128\LS) and of $T_{\rm eff}=5851$\thinspace K (120\LS) in Section~\ref{ssec.chaotic}. It results from the evolution at the proximity of the unstable saddle point.

\begin{figure*}
\centering
\resizebox{\hsize}{!}{\includegraphics{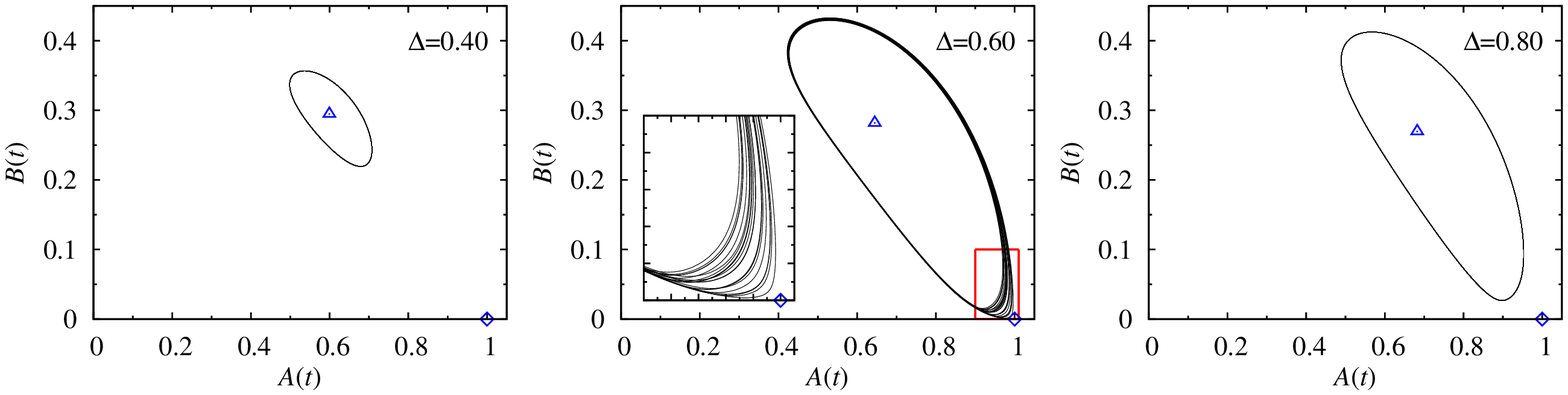}}
\caption{Trajectories in the ($A(t)$, $B(t)$)-plane for several modulation cycles and different $\Delta$. Unstable single-mode and two-mode fixed points are marked with diamond and triangles, respectively.}
\label{fig.ae_aa}
\end{figure*}

\begin{figure}
\centering
\resizebox{\hsize}{!}{\includegraphics{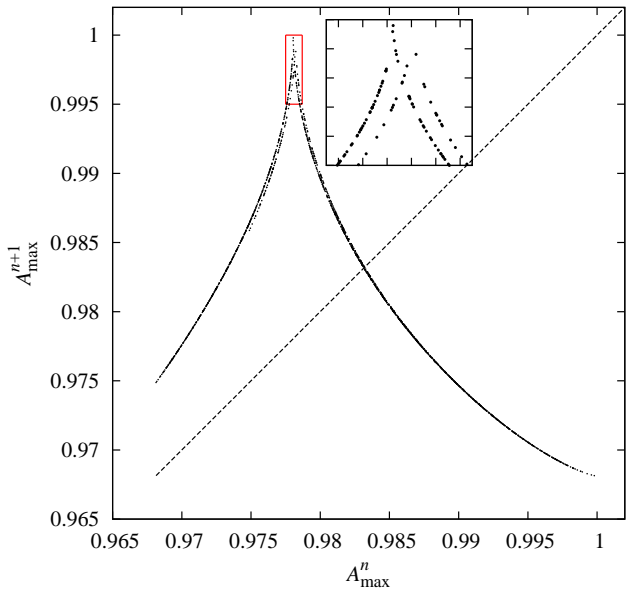}}
\caption{First return map on the successive amplitude maxima of amplitude $A(t)$; $A^{n+1}_{\rm max}$ vs. $A^{n}_{\rm max}$.}
\label{fig.ae_rmAmax}
\end{figure}

To analyse the dynamics of our models in more detail we follow \cite{bk11} and consider the phase plots of amplitudes, i.e. plots of $A(t)$ vs. $B(t)$, which are displayed in Fig.~\ref{fig.ae_aa} for the discussed cases. Note that the ($A$, $B$)-plane is a subset of the full 3D phase-space ($A$, $B$, $\Gamma$) and that now we focus on the long term variation, with pulsations filtered out. For $\Delta=0.4$ and $\Delta=0.8$ the phase plots are very simple -- the trajectory is a single loop and resulting modulation is strictly periodic. In the middle panel of Fig.~\ref{fig.ae_aa}  we plot the trajectories for the most interesting case of $\Delta=0.6$. Clearly, the trajectory is not a single curve now, which is best visible at the proximity of the unstable single-mode point ($A=1$, $B=0$). The phase plot indicates the presence of the strange attractor, which may be further confirmed through construction of the first return map on the consecutive maxima of amplitude $A(t)$. It is plotted in Fig.~\ref{fig.ae_rmAmax} and is very similar to that of the Lorentz attractor \citep[see e.g.][]{owc}, albeit, at close inspection, it shows the double-cusp feature (see inset). The slope of the $A_{\rm max}^{n+1}$ vs. the $A_{\rm max}^n$ at the intersection with diagonal is greater than unity ($|{\rm slope}|>1$) and therefore the behaviour is chaotic. The character of the return map is also very similar to that of \cite{bk11} (see their fig.~3) who studied the 9:2 resonance suspected to cause the Blazhko effect in RR~Lyrae stars. Clearly the two half-integer resonances lead to similar types of dynamical behaviour, either to a strictly periodic modulation, or to a chaotic one. We stress however that our models, both hydrodynamic and based on the integration of AEs are not `too chaotic' -- the modulation may be described as quasi- or more or less periodic. This is exactly what we observe in Blazhko RR~Lyrae stars.

\begin{figure}
\centering
\resizebox{\hsize}{!}{\includegraphics{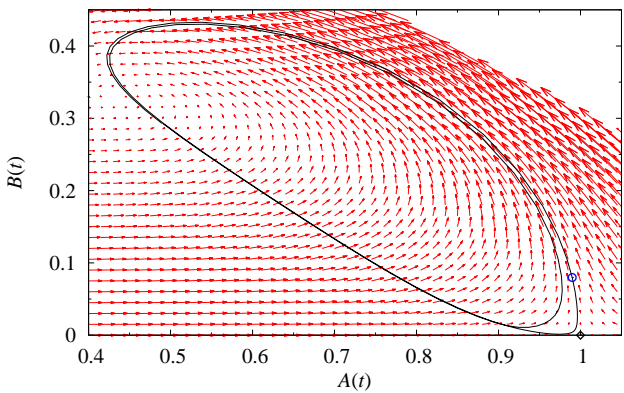}}
\caption{Snapshot from the animation attached as a supplementary online material. Trajectories for two modulation cycles passing either close or farther away the saddle are plotted. The flow field is visualised at the instant marked with the circle.}
\label{fig.ae_flow}
\end{figure}

 The evolution close to the saddle point, in particular the differences in periods of the consecutive modulation cycles, can be understood through analysis of the flow field, i.e. ($\dot{A}$, $\dot{B}$), which varies in time. It is tangent to the trajectory at the point ($A(t)$, $B(t)$). The animation attached as a supplementary on-line material shows the evolution for two trajectories: one located closer to the saddle than the other. The snapshot from this animation is plotted in Fig.~\ref{fig.ae_flow}. It is clear that at the saddle trajectories are at first attracted, but repelled later on. Also the evolution at the saddle is slow (short vectors in the flow field) -- the closer to the saddle the slower the evolution. Thus, the modulation cycles corresponding to the two trajectories have different length.

\begin{figure}
\centering
\resizebox{\hsize}{!}{\includegraphics{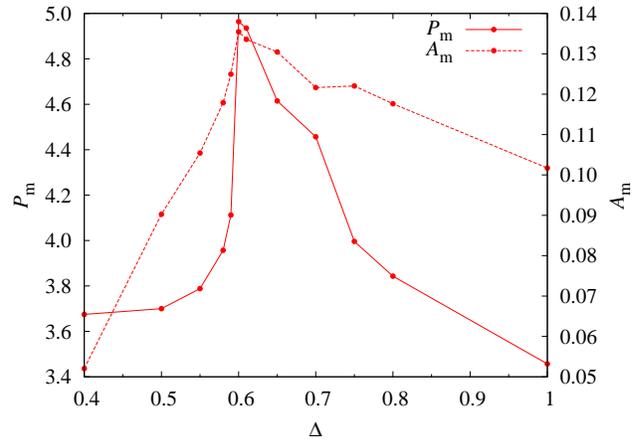}}
\caption{Mean modulation periods and modulation amplitudes as a function of $\Delta$.}
\label{fig.ae_mp}
\end{figure}

In Fig.~\ref{fig.ae_mp} we plot the mean modulation periods and modulation amplitudes, as defined in Section~\ref{ssec.periodic}, for a set of models with different $\Delta$. Just as in the case of the hydrodynamic models (of $120\LS$ and $128\LS$, see Fig.~\ref{fig.hyd.mp}) we observe the increase of the mean modulation period, accompanied with the increase of the modulation amplitude. The longest modulation period is found for $\Delta=0.6$ for which the modulation is chaotic.

\section{Discussion}\label{sec.concl}

\begin{itemize}
\item {\it Origin of the period doubling and of modulation of pulsation.} In Fig.~\ref{fig.hyd.hr} loci of several half-integer resonances are plotted. In principle each of them may cause the period doubling effect, provided that the resonant mode is not damped too strongly, and that respective model falls close to the resonant centre \citep{mb90}. These conditions rule out all the resonances displayed in Fig.~\ref{fig.hyd.hr} except the 3:2 resonance between the fundamental mode and the first overtone. In Fig.~\ref{fig.hyd.orig} we plot the amplitude of the period doubling (amplitude of the highest peak at around $f_0/2$) vs. the mismatch parameter for the three half-integer resonances. Only models of $136\LS$ are plotted as for this luminosity we conducted a model survey through the full period doubling domain. It is clear that only for the 3:2 resonance all the models are close to the resonance centre, with $|\Delta_{3:2}|<0.06$ (where $\Delta_{3:2}=\omega_1/\omega_0-1.5$). Considering e.g. the 7:2 resonance with the fourth overtone (middle panel of Fig.~\ref{fig.hyd.orig}), although some models are located at direct proximity of the resonance centre, other models within the period doubling domain may be located as far as $|\Delta_{7:2}|>0.5$, which rules out this resonance as a possible cause of period doubling. The same holds for the 9:2 resonance with the seventh overtone. 

The domain with the modulation of pulsation lays within the period doubling domain. Its location with respect to the mismatch parameters is also plotted in Fig.~\ref{fig.hyd.orig}, in which we used the amplitude of the highest modulation peak in the vicinity of $f_0$ ($f_0+f_{\rm m}$ or $f_0-f_{\rm m}$) to characterise the strength of modulation. For clarity, only the domain for $L=136\LS$ is plotted, the domains for other luminosities fall roughly at the same location. Already from Fig.~\ref{fig.hyd.hr} it is clear that the domain with modulation of pulsation follows (at lower luminosities) the loci of the 7:2 resonance with the fourth overtone. Indeed, in Fig.~\ref{fig.hyd.orig} the resonance centre falls in the middle of the modulation domain and all of the models lay in close proximity to the resonance centre -- as one may read from Fig.~\ref{fig.hyd.orig}  ($|\Delta_{7:2}|<0.05$ for all the models). This is however true also for the 3:2 resonance that causes the period doubling. Although the models are at offset with respect to the resonance centre, they all fall within $0.02<\Delta_{3.2}<0.04$. As the modulation domain arises within the period doubling domain, which is caused by the 3:2 resonance, it is therefore natural to assume that the same resonance causes the modulation. It is confirmed with the analysis of the amplitude equations presented in Section~\ref{sec.ae} \citep[and also in][]{bk11}. No additional resonance is needed to cause the modulation. It appears within the period doubling domain, not at its centre but at positive $\Delta$ (Fig.~\ref{fig.ae_mp}; note a different definition of $\Delta$ at this plot). Note that the positive mismatch as defined in eq.~(\ref{eq.delta}) in Section~\ref{sec.ae} corresponds to $\Delta_{3:2}>0$, just as we have for our hydromodels (Fig.~\ref{fig.hyd.orig}). Taking into account that all the behaviours we have found in our hydrodynamic models (period doubling, periodic and quasi-periodic modulation) can be reproduced with AEs for the 3:2 resonance only, we conclude that this resonance is the only cause of both period doubling and amplitude modulation in our hydrodynamic models. 

\begin{figure}
\centering
\resizebox{\hsize}{!}{\includegraphics{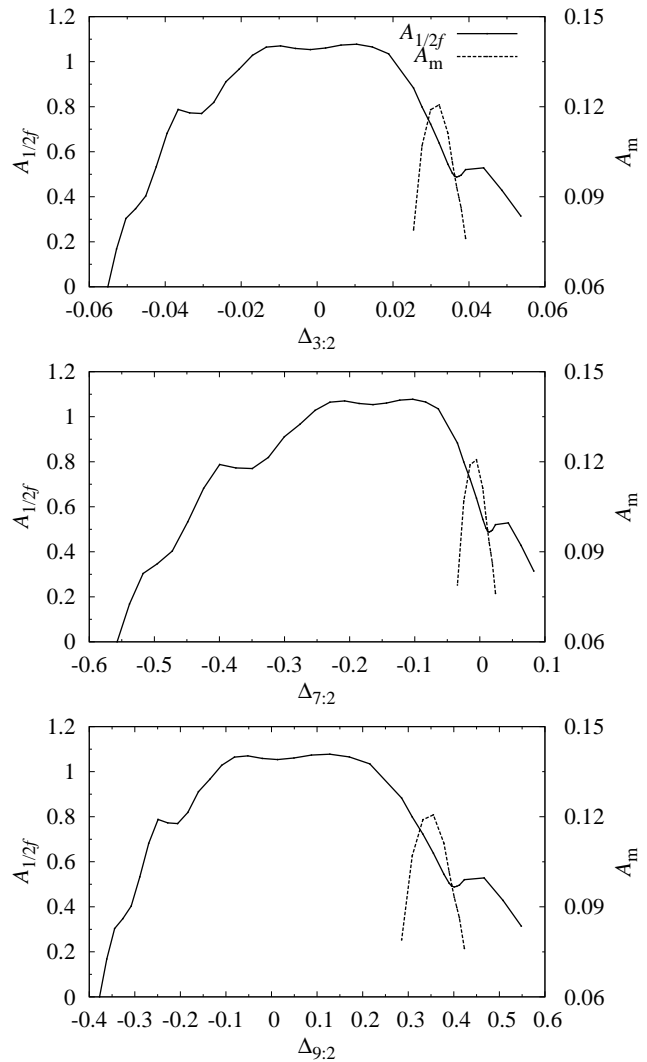}}
\caption{Amplitude of the sub-harmonic frequency, $A_{1/2f}$, and amplitude of the highest modulation peak at $f_0$, $A_{\rm m}$, plotted versus the mismatch parameters for (top panel) 3:2 resonance with the first overtone, $\Delta_{3:2}=\omega_1/\omega_0-1.5$, (middle panel) 7:2 resonance with the fourth overtone, $\Delta_{7:2}=\omega_4/\omega_0-3.5$ and (bottom panel) 9:2 resonance with the seventh overtone, $\Delta_{9:2}=\omega_7/\omega_0-4.5$.}
\label{fig.hyd.orig}
\end{figure}

\item {\it Reliability of the computed models.} Our models have strongly decreased eddy viscosity as compared to the standard models of BL~Her stars, i.e. $\alpha_{\rm m}=0.05$ to be compared with $\alpha_{\rm m}=0.2$ we used in \cite{ssm12}. Consequently the pulsation is violent and spurious spikes appear in the luminosity curve. Still the models may represent the phenomena that may occur in real BL~Her stars. Certainly it is so for the period doubling discovered in these stars only recently, 20 years after the effect was predicted based on hydrodynamic models (see Introduction). No modulation of pulsation was observed in BL~Her stars so far.

The turbulent convection model that we use \citep{ku86} is a very simple 1D formula with several free parameters. The treatment of viscous effects seem to be the weakest point of this and similar models, which is not surprising. The viscous dissipation of turbulent energy takes place over many length scales and its one parameter description is, out of necessity, provisional. We note that the current convective models cannot reproduce a double-mode pulsation unless unphysical neglect of negative buoyancy below the convective zone leads to artificial viscous damping there \citep{sm08b,sm10}. With the purely radiative models double-mode pulsation couldn't be reproduced either, unless the artificial dissipation (replaced in convective codes by viscous dissipation) was decreased \citep{kovb93}. Similarly, the period doubling in RR~Lyrae models of \cite{kms11} was produced assuming lower values of eddy viscosity. Thus, discovery of modulation of pulsation in BL~Her stars in the future cannot be excluded. 

\item {\it Models with chaotic modulation of pulsation.} In case of $L=136\LS$ we computed additional models across the instability strip. This allowed us to detect another domain in which pulsation is modulated, this time however, in a very chaotic manner. In these models we observe a wealth of dynamical behaviours. In particular we identify several stability windows within the chaotic regime, with stable period $k$ cycles (e.g. with $k=3$, $k=7$ or $k=9$). They arise due to the tangent bifurcation and, as the control parameter ($T_{\rm eff}$) is increased, undergo a series of pitchfork bifurcations (sub-harmonic cascade). These models will be subject of a separate paper (Smolec \& Moskalik, in prep.).
\end{itemize}

\section{Implications for the Blazhko effect in RR~Lyrae stars}\label{sec.impl}

The modulation of pulsation is a common property of RR~Lyrae stars, where it is called the Blazhko effect. In several stars period doubling was also detected and was reproduced by hydrodynamic RR~Lyrae models of \cite{kms11}, which however, show no trace of modulation. Period doubling in the hydrodynamic RR~Lyrae models is caused by the 9:2 resonance between the fundamental mode and the ninth overtone. The analysis of amplitude equations by \cite{bk11} shows that the same resonance may cause the modulation of pulsation. Confirmation of this result through direct hydrodynamic modelling would strongly support this radial resonance model. Our hydrodynamic BL~Her models are the first models that demonstrate that the mechanism may indeed work. Except for larger luminosities, BL~Her stars are siblings of RR~Lyrae stars. The resonance that causes the modulation of pulsation in our models, 3:2 resonance with the first overtone, cannot work in case of RR~Lyrae stars. Its locus falls beyond the RR~Lyrae instability strip. It is clear however, that the underlying dynamics is the same for the 3:2 and for the 9:2 resonances as studies of amplitude equations indicate \citep[Section~\ref{sec.ae} and][]{bk11}. 

The great advantage of the radial resonance model is that it can cause the irregular modulation, observed in Blazhko RR~Lyrae stars. The striking evidence comes from the nearly continuous observations of the satellite missions \citep[see e.g.][]{bea10,eg12}. Our models together with analysis of amplitude equations demonstrate how these irregularities may arise in the framework of the resonance model. For some model parameters the modulation is no longer strictly periodic, but becomes chaotic. The models indicate that it happens when amplitudes of the involved fundamental and first overtone modes approach the unstable saddle point which corresponds to single periodic fundamental mode pulsation. Then the period doubling is quenched and the variation of the modal amplitudes is very slow: the closer to the saddle, the slower the modulation. Thus consecutive modulation cycles differ. It is worth noting that our models showing the irregular modulation of pulsation reproduce other features of Blazhko RR~Lyrae stars that simultaneously show the period doubling effect. We observe qualitatively the same properties of Fourier spectra at half integer frequencies -- bunches of peaks, indicating the irregular nature of period doubling phenomenon, as well as swaping of higher/lower maximum cycles during the modulation \citep[][Moln\'ar et al., submitted]{szabo10, kol11}.

Comparison of the properties of modulation in our BL~Her models to that observed in the Blazhko RR~Lyrae stars points to the possible challenges for the radial resonance model. One of them is a very strong period doubling effect in all our models. On the contrary, only in a few Blazhko stars observed with satellite missions the effect was detected and its amplitude is rather low \citep{szabo10}. The strong period doubling is also evident in our models based on the amplitude equations integration, as well as in models corresponding to the 9:2 resonance analysed by \cite{bk11}, which should be operational in RR~Lyrae stars. The lack of period doubling in most of the Blazhko stars indicate that the modulation affects predominantly the fundamental mode, while the amplitude of suspected resonant mode must remain quenched during most of the modulation cycle. Our models indicate that the period doubling is quenched while the model evolves close to the single-mode saddle point, where the amplitude of the resonant mode approaches to zero. The resonant mode will always be modulated in the resonant scenario -- the modal amplitudes must encircle the unstable two mode fixed point -- see Fig.~\ref{fig.ae_aa}. To decrease the modulation of the resonant mode and hence to quench the period doubling, one has to bring the two-mode fixed point closer to the $B(t)=0$ axis. An extensive parameter study of the amplitude equations is needed to point under which conditions such modulation is possible.

Irregularities are often detected in the Blazhko stars. In our models they seem to be restricted to a very narrow parameter range, most of our models show a strictly periodic modulation. This might not be a problem for the resonant scenario if one takes into account the dispersion of stellar parameters (masses, luminosities, metallicities) of real stars or if one allows other resonances to play a dynamical role. The latter possibility is worth considering. Nearly 50 per cent of RR~Lyrae stars (pulsating in the fundamental mode) show the Blazhko effect \citep{jj09,bea10}. It seems unlikely that the same, high-order resonance operates in all these stars. All other resonances within the radial modes face the same problems (or even harder, as most of the overtones are heavilly damped). The half-integer resonances with non-radial modes may present a solution. Non-radial modes are densely packed in between the radial modes, thus there is always a candidate for resonant coupling. On the other hand, the mode selection would be a crucial problem here. We note that in recent years non-radial modes were detected in several RR~Lyrae stars, both in the ground-based and space observations \citep[e.g.][]{grub,om09,eg12}.

\section*{Acknowledgments}
We are grateful to Katrien Kolenberg for reading and commenting the manuscript. This research has been supprted by Polish National Science Centre (DEC-2011/01/M/ST9/05914). Model computations presented in this paper were conducted on the psk computer cluster in the Copernicus Centre, Warsaw, Poland.

\section*{Supporting Information}
Additional Supporting Information may be found in the online version of this article:

\smallskip
\noindent{\bf Animation.} Model's evolution in the $B(t)$ vs. $A(t)$ plane ($\Delta=0.6$) for the two modulation cycles. Diamond marks the location of unstable single-mode fixed point (saddle). Circle shows the model's position during the evolution. The instantaneous flow field ($\dot{A}(t)$, $\dot{B}(t)$) is plotted with vectors. Snapshot from the animation is plotted in Fig.~\ref{fig.ae_flow}.

\label{lastpage}

\end{document}